\documentclass[twocolumn,showpacs,preprintnumbers,amssymb]{revtex4-1}

\usepackage{latexsym,amsfonts,amsmath,amssymb,graphics}
\usepackage{dsfont}
\usepackage[dvips]{graphicx}
\usepackage{mathrsfs}  
\usepackage{bbm}       
\usepackage{slashed}   
\DeclareMathAlphabet{\mathpzc}{OT1}{pzc}{m}{it}

\newcommand{\refer}[1]{(\ref{#1})}

\def\nn{\nonumber}
\def\bea{\begin{eqnarray}}
\def\eea{\end{eqnarray}}
\def\beq{\begin{equation}}
\def\eeq{\end{equation}}

\newcommand{\tr}[0]{{\rm tr}}

\newcommand{\KR}[0]{\mathcal{R}}

\newcommand{\tig}[0]{{\widetilde{\Gamma}}}

\newcommand{\OLB}[0]{{(1B)}}

\newcommand{\anti}[1]{\overline{#1}}

\newcommand{\volfour}[1]{\!\! \frac{{\rm d}^4 #1}{\left(2\pi\right)^4}}

\newcommand{\dd}[0]{{\rm d}}

\newcommand{\cO}[0]{{\mathcal{O}}}

\newcommand{\al}{\alpha}

\newcommand{\ga}{\gamma}
\newcommand{\Ga}{\Gamma}

\newcommand{\La}{\Lambda}

\begin{document}
\title{Ultraviolet cutoffs and the photon mass}

\author{Piotr H. Chankowski$^1$, Adrian Lewandowski$^{2,1}$,
Krzysztof A. Meissner$^{1}$}
\affiliation{$^1$ Faculty of Physics,
University of Warsaw\\
Pasteura 5, 02-093 Warsaw, Poland\\
$^2$ Max-Planck-Institut f\"ur Gravitationsphysik
(Albert-Einstein-Institut)\\
M\"uhlenberg 1, D-14476 Potsdam, Germany\\
}

\begin{abstract}
{\bf Abstract}
The momentum UV cutoff in Quantum Field Theory is usually treated as an 
auxiliary device allowing to obtain finite amplitudes satisfying all physical 
requirements. It is even absent (not explicit) in the most popular approach 
- the dimensional regularization. We point out that the momentum cutoff 
treated as a bona fide physical scale, presumably equal or related to the 
Planck scale, would lead to unacceptable predictions. One of the dangers is 
a non-zero mass of the photon. In the naive approach, even with the cutoff 
equal to the Planck scale, this mass would grossly exceed the existing 
experimental bounds. We present the actual calculation using a concrete 
realization of the physical cutoff and speculate about the way to restore 
gauge symmetry order by order in the inverse powers of the cutoff scale.  
 
\end{abstract}
\maketitle

In usual applications of Quantum Field Theory (QFT) the momentum cutoff 
(explicit or, as in the Dimensional Regularization, implicit) is treated 
as an auxiliary parameter and sent to infinity at the end of the 
renormalization procedure. However in the context of a quest for a 
fundamental theory unifying elementary particle interactions with gravity, 
QFT models should be viewed as only effective theories with a real momentum 
cutoff which, as in QFT applications to statistical physics problems, should 
have a concrete physical interpretation, most probably of the intrinsic
scale $\La$ of the underlying fundamental theory. In this short note, based 
on the previous work \cite{CLMN} 
and on the accompanying paper \cite{ZJlong} (where all the relevant 
references can be found) we would like to point out some important aspects
of treating the cutoff scale as a {\it bona fide} physical scale $\La$ (the 
problem was partly analyzed in connection with quadratic divergences 
in QFT \cite{Fujikawa,EJ,MN}). 

The most spectacular danger of keeping $\Lambda$ finite is, unless the 
effective field theory is of very special form, generation of the photon 
mass proportional to inverse powers of $\La$. This is because the gauge 
symmetry ensuring the vanishing of the photon mass for $\La\to\infty$, for 
finite $\La$ remains generically broken. Since the bounds are extremely 
stringent, even the natural assumption $\La\approx M_{Pl}$ ($M_{Pl}$ being
the Planck scale) could lead to 
unacceptably large photon mass, bigger than the experimental limit. In this 
note we illustrate this on a simple example and speculate how the problem 
could possibly be avoided in the context of an underlying more fundamental 
finite theory.

To define the framework we consider first renormalization of a general
YM theory choosing (out of many other possibilities) the momentum
cutoff regularization which consists of modifying {\it every} derivative 
in the Lagrangian (including the recursively generated counterterms - see
below)
according to the rule
\bea
\partial_\mu\rightarrow\exp(\partial^2/2\Lambda^2)\partial_\mu~\!.
\label{eqn:PrescriptionDef}
\eea
In the momentum space this prescription corresponds to the replacement
\bea
k_\mu\rightarrow{\cal R}_\mu(k)\equiv\exp(-k^2/2\Lambda^2)~\!k_\mu~\!.
\label{eqn:MomSpacePrescription}
\eea
For instance, the regularized ghost contribution to the vacuum polarization 
tensor  (diagram $C$ in Fig. 1) reads
\beq
\tig^{\mu\nu}_{\alpha\beta}(l)=-\tr(e_\alpha e_\beta)
\int\volfour{k}\, i\, \frac{\KR^\mu(k)\KR^\nu(k+l)}{\KR^2(k)\KR^2(k+l)}~\!.
\eeq
where $e_\alpha$ are the antihermitian generators of the adjoint 
representation with included coupling constants (i.e. 
$e_\alpha= g\, T_{\alpha}^{ADJ}$ for a simple gauge group).

With the replacement \refer{eqn:MomSpacePrescription} the Wick rotation is, 
strictly speaking, not justified and neglecting the integral over the contour 
at infinity must be regarded a part of the regularization prescription 
(alternatively, the prescription can be formulated directly in the Euclidean 
version of the theory). 

As the standard analysis carried out in \cite{ZJlong} shows, in the 
regularization \refer{eqn:PrescriptionDef} all diagrams of a renormalizable 
theory are convergent with the exception of one-loop vacuum graphs (which 
anyway cannot appear in physically interesting amplitudes as divergent 
subdiagrams). Computation of diagrams regularized in this way is more 
complicated than in the Dimensional Regularization but still manageable. For 
example, each one-loop diagram can be expressed in terms of the confluent 
hypergeometric function
\beq
U\!\left(a,b,z\right)=\frac{1}{\Gamma(a)}\int\limits_0^\infty \!\dd t~\!
t^{a-1}(1+t)^{b-a-1}\exp(-z t)~\!,\nn
\eeq
after applying the expansion
\beq
\nn
{1\over{\cal R}^2(k)-m^2}={e^{k^2/\Lambda^2} \over k^2-m^2}\sum_{n=0}^\infty
\left[{m^2\over m^2-k^2}\left(1-e^{k^2/\Lambda^2}\right)\right]^n,
\eeq
to all regularized propagators.
The first term in this expansion bears a very close resemblance to the 
propagator used in the context of Wilson's exact renormalization group 
equations \cite{BallThorne}. The advantage of our expression is that it 
is better suited for theories with spontaneous symmetry breaking, in which 
$m^2$ in general depends on background scalar fields (which can
keep track of vacuum expectation values of dynamical scalar fields). In the 
Euclidean space, for $k^2\rightarrow-k^2_E$, the above series is absolutely 
convergent. In particular, owing to the growing powers of $m^2-k^2$ in 
successive terms, only a finite number of terms of a given one-loop diagram  
yield integrals which are divergent when the factors 
$e^{k^2/\Lambda^2}(1-e^{k^2/\Lambda^2})^n$ are omitted. The remaining terms would 
be integrable without these factors which implies that their contributions 
vanish in the limit $\Lambda\rightarrow\infty$. 

Since the regularization \refer{eqn:PrescriptionDef} breaks the gauge (more
precisely, the BRST) invariance, a special subtraction scheme in necessary
in order to arrive at a finite (renormalized) one-particle irreducible
effective action $\Gamma$ (the generator of the strongly connected Green's 
functions) satisfying \emph{in the limit $\La\rightarrow\infty$} the 
requirements of the BRST invariance (embodied in the appropriate functional 
identity). In the accompanying paper \cite{ZJlong} such a subtraction 
scheme (called $\La$-$\anti{\rm MS}$), belonging to the class of mass 
independent schemes and adapted to the regularization 
\refer{eqn:PrescriptionDef}, is proposed. It relies on the Quantum Action
Principle \cite{QAP} and is defined 
recursively in the following way. Having a local action $I_n^\Lambda$ (with 
all counterterms up to the order $\hbar^n$ included and with the replacement
\refer{eqn:PrescriptionDef} made in all derivatives), one considers $\Ga_n$ - 
the asymptotic part of $\Ga^\La_n\equiv\Gamma[I_n^\Lambda]$, obtained 
from it by omitting all terms that would vanish for $\Lambda\to\infty$ and
constructs order $\hbar^{n+1}$ ``minimal'' counterterms 
$-\Gamma_n^{(n+1) {\rm div}}$ which subtract 
(appropriately defined) ``pure'', order $\hbar^{n+1}$, divergences of $\Ga_n$. 
In the next step one constructs finite non-minimal counterterms 
$\delta_\flat\!\Gamma_n^{(n+1)}$ of the restricted
schematic form ($A^\mu$, $\phi$ and $\psi$ stand respectively for generic 
gauge, scalar and fermion fields)
\bea
\label{Eq:Non-minimal-ct}
\delta_\flat\!\Gamma_n^{(n+1)}&\in&\int
(\partial^\mu\! A_\mu)(\partial^\nu\! A_\nu)\oplus A_\mu A^\mu\oplus
A_\mu\anti{\psi}\gamma^\mu P_L\psi\nn \\ &{}& \hspace*{-40 pt} 
 \oplus A_\mu\anti{\psi}\gamma^\mu P_R\psi\oplus
\phi\phi A_\mu A^\mu\oplus A_\mu\partial^\mu\!\phi\oplus
\phi A_\mu\partial^\mu\!\phi
\nn \\ &{}&  \hspace*{-40 pt}
\oplus \phi A_\mu A^\mu\oplus A A\,\partial\! A\oplus AAAA~\!,\quad
\eea
so that $I_{n+1}\equiv I_n-\Gamma_n^{(n+1) {\rm div}}+\delta_\flat\!\Gamma_n^{(n+1)}$
leads to $\Gamma_{n+1}$ - the asymptotic part of 
$\Ga^\La_{n+1}\equiv\Gamma[I_{n+1}^\Lambda]$ - which up to terms of 
order $\hbar^{n+1}$ is finite and satisfies the Slavnov-Taylor 
identities (STIds) following from the required BRST invariance.
In \cite{ZJlong} it is shown that the choice \refer{Eq:Non-minimal-ct},
which is particularly natural (no non-minimal counterterms are generated for terms of the action without gauge fields), satisfies all the requirements and is equivalent to the
usual specification of the renormalization conditions. As a result of the 
procedure sketched above one constructs the action $I_\infty$
\begin{eqnarray}
I_\infty=I_0+\sum_{n=0}^\infty(-\Gamma_n^{(n+1) {\rm div}}+\delta_\flat\!\Gamma_n^{(n+1)})
\label{eqn:Iinfty}
\end{eqnarray}
expressed in terms of renormalized parameters and couplings, depending
explicitly on $\Lambda$ (through the counterterms $-\Gamma_n^{(n+1) {\rm div}}$) and 
such that Green's functions obtained from $\Gamma[I_\infty^\Lambda]$ satisfy 
STIds \emph{in the strict limit $\Lambda\rightarrow\infty$}.

Being mass independent, the proposed scheme introduces, similarly as the
ordinary $\overline{\rm MS}$ scheme, an auxiliary mass scale $\mu$. We have 
verified by explicit one-loop calculations in a general renormalizable model 
(with a non-anomalous fermionic representation) that the proposed subtraction 
scheme is equivalent to the standard $\anti{\rm MS}$ scheme based on the
dimensional regularization (DimReg) with the anticommuting $\gamma^5$ matrix
(the so-called naive prescription): the one-loop 1PI effective action in 
$\Lambda$-$\anti{\rm MS}$ can be obtained from its $\anti{\rm MS}$ 
counterpart via a reparametrization (a ``finite renormalization") of fields 
and couplings. Furthermore, we have proved recursively, that the finite 
effective action $\Gamma[I^\Lambda_\infty]$ and the action $I^\Lambda_\infty$ 
itself satisfy the Renormalization Group Equations (RGE) which ensure 
independence of physical result of the auxiliary mass scale $\mu$. (The  
one-loop equivalence of the $\anti{\rm MS}$ and $\Lambda$-$\anti{\rm MS}$ 
schemes allowed us to obtain in \cite{ZJlong} the two-loop RGE for the 
$\Lambda$-$\anti{\rm MS}$ scheme parameters). Finally, we have performed 
some two-loop consistency checks as well.   

Established RG invariance of $I^\La_\infty$ consisting of the regularized 
original action $I^\La_0$ and the constructed counterterms, in which the 
replacement \refer{eqn:PrescriptionDef} is made (as pointed out in 
\cite{ZJlong}, this is necessary for consistency of the entire scheme),
allowed to show that, despite not having the same functional form as $I_0$ 
(e.g. each interaction term depending on the gauge fields $A^\mu$ is 
multiplied by a different series of renormalized couplings with coefficients 
divergent as $\Lambda\rightarrow\infty$), it does wind up into a bare action 
$I_{\rm B}$ which depends on $\Lambda$ only through the appropriately defined 
bare parameters and through the regularizing exponential factors 
\refer{eqn:PrescriptionDef} accompanying derivatives. This (technically 
nontrivial in the case of a gauge symmetry violating regularization) result 
opens the possibility to view $I^\La_\infty$ (after expanding the exponentials, 
so that they give rise to infinite sum of higher and higher dimension 
operators) as a part of the complete Lagrangian density of an effective 
field theory which in the perturbative expansion reproduces results of 
some {\it finite} fundamental theory of all interactions. The scale 
$\Lambda$ should be then identified with an intrinsic physical scale of 
the putative fundamental theory rather than with the scale introduced by 
the Wilsonian procedure of integrating out some high energy degrees of 
freedom. For this interpretation to be possible it is, however, 
indispensable to address the problem of the residual breaking of the gauge 
(BRST) invariance by terms suppressed by inverse powers of $\Lambda$ of which
one of the consequences is the photon mass generation.

To illustrate the problem 
we consider 
here, using the regularization \refer{eqn:MomSpacePrescription}, the one-loop 
contribution to the standard gauge field self energy (vacuum polarization) 
tensor $\tig^{\mu\nu}_{\alpha\beta}(l)$ contracted with the four-momentum $l_\mu$. 
Before making subtractions (as indicated by the superscript $1B$) we find 
(in the Landau gauge, using a developed Mathematica package described in 
\cite{ZJlong}): 
\begin{eqnarray}
l_\mu \tig^{\mu\nu}_{\alpha\beta}(l)^\OLB&=&
-3g_s^2\,\delta_{\alpha\beta}\,\frac{l^\nu}{(4\pi)^2}
\big\{-{\Lambda^2}-\frac{5}{24}l^2\nonumber\\ 
&{}& \nn -{7\over384}\frac{l^4}{\Lambda^2}
+{1\over1536}\frac{l^6}{\Lambda^4}+\cO(\La^{-6})\big\}\,,
\end{eqnarray}
in ``QCD without quarks" (diagrams $A$, $B$ and $C$ in Fig. 1), and
\begin{eqnarray}
l_\mu \tig^{\mu\nu}_{}(l)^\OLB&=&e^2\,\frac{l^\nu}{(4\pi)^2}
\left\{-\frac{2}{3}(l^2-3m^2+3\La^2)\right.\nonumber\\   
&{}& \hspace*{-120 pt} 
\phantom{e^2\,\frac{l^\nu}{(4\pi)^2}\left\{\right.}
\left.+{1\over\La^2}\left[
-{11\over96}l^4+l^2 m^2+6m^4 \ln{m^2\over(0.37 \La)^2}\right]
+\cO(\La^{-4})\right\},\nonumber
\end{eqnarray}
in QED with a single charged lepton of mass $m$ (diagram G). These 
contributions clearly give non-vanishing correction to the gluon and photon 
masses, respectively.

Terms that survive in the limit $\La\to\infty$ clearly show that the gauge 
symmetry is badly broken by the regularization prescription 
\refer{eqn:PrescriptionDef}. As explained above, they are removed by 
local counterterms (i.e. minimal subtraction of the divergent part followed 
by the addition of the first two terms in \refer{Eq:Non-minimal-ct} with 
appropriate coefficients). On the other hand, the terms of the above 
expressions suppressed by the inverse powers of $\La$ also break the gauge
invariance but are not subtracted by adding non-minimal counterterms which 
are (in the construction following from the QAP principle) independent of 
$\Lambda$. If present, they would imply a contribution to the photon mass 
$m_\ga$ of the order of $(\al_{\rm EM}/4\pi)^{1/2} M_{\rm top}^2/\Lambda$. 
In the usual 
approach ($\La\to\infty$) such terms would be absent but if $\La$ is a physical 
cutoff scale we have to consider them as possible genuine corrections. 
However, because of the experimental limit on the photon mass 
($m_\gamma<1.7\times 10^{-22}$ GeV from the dispersion relations from pulsar 
emissions \cite{BW} and $m_\gamma< 10^{-27}$ GeV from combination of all
data \cite{PDG}) this kind of breaking is excluded even for $\Lambda$ as 
high as the Planck scale which would give
\beq
|m_\ga|\approx \left({\al_{\rm EM}\over4\pi}\right)^{1/2}{M_{top}^2\over M_{Pl}}
\approx 10^{-18}~{\rm GeV}.\nn
\eeq
In fact, the situation is even worse since $m_\ga^2$ generated in the above 
example not only grossly exceeds the experimental limits but has also the 
wrong sign. 

Therefore, if the cutoff $\La$ is to be treated as a \emph{finite} physical 
scale of an underlying fundamental theory, one has to assume that the 
\emph{complete} bare action $I^{\rm complete}_{\rm B}$ of the effective QFT, which 
reproduces all results (including those depending on the gravitational sector) 
of the latter theory has also additional, as compared to the local action 
$I_{\rm B}=I_\infty^\Lambda$ (counter)terms suppressed by inverse powers of 
$\Lambda$ which conspire to restore exact BRST invariance of the amplitudes. 
The structure of the residual gauge symmetry breaking revealed by the above 
two examples suggests that such a solution may be viable: the $\Lambda$ 
suppressed terms which must be subtracted do not involve logarithms of 
momenta so that $I^{\rm complete}_{\rm B}$ can still be analytic. Additional 
terms with higher derivatives which must be present in $I^{\rm complete}_{\rm B}$ 
would complement those which implementing the regularization 
\refer{eqn:MomSpacePrescription} reflect finiteness of the underlying theory - 
the assumption that higher derivative terms of $I^{\rm complete}_{\rm B}$ combine 
solely to exponential factors \refer{eqn:PrescriptionDef} of an otherwise 
\emph{renormalizable} action is certainly too simplistic for the action of 
an effective field theory corresponding to the fundamental theory of all 
interactions. In turn the presence of logarithms of  masses (which in 
general dependent on the background fields), simply indicates that in order 
to restore BRST-invariance of all amplitudes for finite $\La$, the 
$I^{\rm complete}_{\rm B}$ should depend on non-polynomial functions of fields
(similarly as the action of the field theory describing the Ising lattice 
model depends on cosh$(\phi)$, where $\phi$ is the order parameter field), 
which after expansion around the background (vacuum expectation values) give 
rise to vertices with arbitrary numbers of scalar fields.

The ultimate structure of $I^{\rm complete}_{\rm B}$ would therefore be such 
as can naturally be expected on the basis of the general principle of
 constructing effective theories.  In the case considered here
it is tempting to assume that the limit $\La\rightarrow\infty$ corresponds
in the effective theory to complete neglect of a gravitational sector,
which for finite $\Lambda$ is entangled with the other sectors and is
indispensable for consistency.

\vspace*{0.2cm}
Summarizing, we conjecture that even if the cutoff scale $\La$ is a real 
physical scale it is possible to introduce local counterterms to the bare 
action that restore the requisite identities and render the vanishing of 
the photon mass order by order in the inverse powers of the scale $\La$.

\vspace{0.2cm}
\noindent{\bf Acknowledgments:} We thank H. Nicolai for discussions.
K.A.M. thanks the Albert Einstein Institute 
in Potsdam for hospitality and support during this work. A.L. and K.A.M. 
were supported by the Polish NCN grant DEC-2013/11/B/ST2/04046.

\onecolumngrid
\vspace{0.9 cm}
\begin{figure}[ht]
\hspace{0.0cm}
\centering
\includegraphics[scale=0.8]{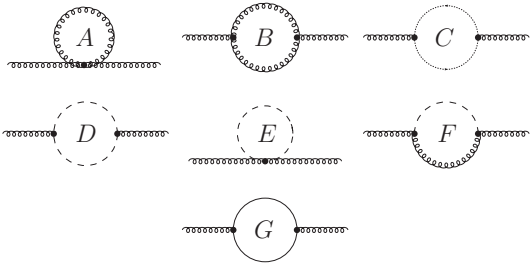}
\label{fig:figure1}
\caption{One-loop contributions to the vacuum polarization. 
}
\end{figure}


\begin{thebibliography}{99}

\bibitem{CLMN} P.H.~Chankowski, A.~Lewandowski, K.A.~Meissner and H.~Nicolai,
  Mod.Phys.Lett. A{\bf30} (2015) no.02, 1550006. 

\bibitem{ZJlong}  P.~H.~Chankowski, A.~Lewandowski and K.~A.~Meissner, arXiv:1608.02270 [hep-ph].


\bibitem{EJ} M.B.~Einhorn and D.R.T.~Jones,
  Phys.\ Rev.\  D{\bf46} (1992) 5206.

\bibitem{MN} K.A.~Meissner and H.~Nicolai,	
  Phys.\ Lett.\ B{\bf660} (2008) 260. 

\bibitem{Fujikawa} K.~Fujikawa,
  Phys.\ Rev.\ D{\bf83} (2011) 105012.

\bibitem{BallThorne} R.~D.~Ball and R.~S.~Thorne,
  Annals of Phys.\  {\bf 236} (1994) 117,
  {\bf 241} (1995) 337.

\bibitem{QAP} Y.~M.~P.~Lam,
  Phys.\ Rev.\ D{\bf 6} (1972) 2145; Phys.\ Rev.\ D{\bf 7} (1973) 2943;
  J.~H.~Lowenstein, Phys.\ Rev.\ D{\bf 4} (1971) 2281,
  Commun.\ Math.\ Phys.\  {\bf 24} (1971) 1,

\bibitem{BW} Z. Bay and J. White, Phys. Rev. {\bf D5} (1972) 796.

\bibitem{PDG} K.A. Olive et al. (Particle Data Group), Chin. Phys. C{\bf38}, 
  090001 (2014) and 2015 update.

\end{thebibliography}
\end{document}